\documentclass[12pt,oneside,a4paper]{article}

\usepackage{hyperref}
\hypersetup{colorlinks,bookmarksopen,bookmarksnumbered,citecolor=blue,
linkcolor=black,pdfstartview=FitH,urlcolor=blue}

\usepackage{graphicx}
\usepackage{amssymb}
\usepackage{cite}
\usepackage{bm}
\usepackage{indentfirst}
\usepackage{amsmath}
\usepackage{hhline}
\usepackage{multirow}
\usepackage{color}
\usepackage{here}

\allowdisplaybreaks
\numberwithin{equation}{section}

\definecolor{dred}{rgb}{0.7,0.15,0.09}
\definecolor{dblue}{rgb}{0,0.12,0.64}
\definecolor{dgreen}{rgb}{0.2,0.51,0.19}

\begin{document}

\begin{titlepage}

\begin{flushright}
KANAZAWA-23-10
\end{flushright}

\begin{center}

\vspace{1cm}
{\large\textbf{
Simultaneous detection of boosted dark matter and neutrinos from the semi-annihilation at DUNE
}
 }
\vspace{1cm}

\renewcommand{\thefootnote}{\fnsymbol{footnote}}
Mayumi Aoki$^{1}$\footnote[1]{mayumi.aoki@staff.kanazawa-u.ac.jp}
and 
Takashi Toma$^{1,2}$\footnote[2]{toma@staff.kanazawa-u.ac.jp}
\vspace{5mm}

\textit{
 $^1${Institute for Theoretical Physics, Kanazawa University, Kanazawa 920-1192, Japan}\\
 $^2${Institute of Liberal Arts and Science, Kanazawa University, Kanazawa 920-1192, Japan}
}

\vspace{8mm}

\abstract{
Dark matter direct detection experiments impose the strong bounds on thermal dark matter scenarios. 
The bound can naturally be evaded if the cross section is momentum transfer or velocity dependent. 
One can test such thermal dark matter scenarios if dark matter particles are boosted by some mechanism. 
In this work, we consider a specific semi-annihilation $\chi\chi\to \nu\overline{\chi}$ where $\chi$ ($\overline{\chi}$) is dark matter (anti-dark matter), 
and search for simultaneous detection of the neutrino and the boosted dark matter in the final state at DUNE. 
We find that the energies of the neutrino and boosted dark matter are reconstructed by kinematics. 
In addition, we find that both signals can be testable at DUNE if the dark matter mass is below $8~\mathrm{GeV}$, and the scattering cross section is momentum transfer dependent. 
Even for larger dark matter masses, the two signals can be tested by combination of DUNE and the other experiments such as IceCube/DeepCore and Hyper-Kamiokande. 
}

\end{center}
\end{titlepage}

\renewcommand{\thefootnote}{\arabic{footnote}}
\newcommand{\bhline}[1]{\noalign{\hrule height #1}}
\newcommand{\bvline}[1]{\vrule width #1}

\setcounter{footnote}{0}

\setcounter{page}{1}

\section{Introduction}
Dark matter existing in our universe is being searched by various ways. 
In particular, direct detection experiments exploring dark matter scatterings with various nuclei such as xenon, argon, germanium etc, have high sensitivities. 
However no dark matter signal is found so far, and the resultant bounds on dark matter--nucleon scattering cross sections are extremely strong~\cite{XENON:2018voc, PandaX-4T:2021bab, LZ:2022ufs}. 
Thus these experiments are starting to rule out thermal dark matter scenarios, which are one of primary dark matter candidates. 
If one still considers thermal dark matter scenarios, some mechanism is required to suppress scatterings between dark matter and nucleons 
such as the pseudo-Nambu-Goldstone dark matter~\cite{Gross:2017dan} and pseudo-scalar interacting fermionic dark matter~\cite{Ipek:2014gua}. 
In these scenarios, the scattering amplitude is suppressed by a small momentum transfer (or dark matter velocity) while annihilations into normal matter particles are unsuppressed, 
thus the annihilation cross section should be the canonical value $\sim10^{-26}~\mathrm{cm^3/s}$ for thermal dark matter abundance. 
However it conversely implies that exploring these kinds of models through usual direct detection experiments is extremely difficult. 

We still have chance to explore these models if the dark matter particles are boosted by some mechanism such as scatterings with high energy cosmic rays~\cite{Ipek:2014gua}, 
vacuum decay~\cite{Cline:2023hfw}, 
heavier dark matter annihilations into lighter ones in multi-component dark matter models~\cite{Aoki:2012ub, Aoki:2013gzs, Kong:2014mia, Agashe:2014yua, Aoki:2014lha, Kopp:2015bfa, Alhazmi:2016qcs, Aoki:2017eqn, Aoki:2018gjf, Kim:2019had} and semi-annihilations~\cite{Berger:2019ttc, Berger:2014sqa, Kelly:2019wow, Toma:2021vlw, Guo:2023kqt}. 
Since the interaction of thermal dark matter itself does not have to be very small to avoid the strong bound of the direct detection experiments, 
one can search the boosted dark matter at large volume neutrino detectors such as IceCube/DeepCore~\cite{IceCube:2015rnn}, Super-Kamiokande~\cite{Super-Kamiokande:2015xms}, Hyper-Kamiokande~\cite{Hyper-Kamiokande:2018ofw} 
and Deep Underground Neutrino Experiment (DUNE)~\cite{DUNE:2020ypp}. 

In this work, we consider the boosted dark matter produced by the semi-annihilation process $\chi\chi\to\nu\overline{\chi}$~\cite{Toma:2021vlw}. 
Here we have two kinds of signals consisting of the high energy neutrino and the boosted anti-dark matter whose energies are fixed at $E_{\nu}=3m_\chi/4$ and 
$E_{\overline{\chi}}=5m_\chi/4$ by the kinematics assuming non-relativistic dark matter in the initial state. 
Therefore these particle productions can be distinctive signals of the semi-annihilation process. 
Although we mainly focus on the case that the scattering cross section with a nucleon is proportional to the momentum transfer $Q^{2n}~(n=1,2)$, 
we also consider a constant cross section ($n=0$) for comparison. 
Then we estimate the number of events for these neutrino/boosted dark matter signals from the Sun, 
and atmospheric neutrino backgrounds assuming the DUNE detector because it has advantage for the angular resolution and tracking compared to the other experiments. 
In addition, note that exploring the boosted dark matter with a mild boost factor as our case is difficult at Super-Kamiokande/Hyper-Kamiokande and 
IceCube/DeepCore because the boost factor of dark matter produced by the semi-annihilation process is below the experimental Cherenkov thresholds: $\gamma_\mathrm{Cherenkov}=1.51$ (water) and $1.55$ (ice)~\cite{Agashe:2014yua}.
For the purpose of this work, we use the neutrino event generator \texttt{GENIE} in order to produce the neutrino scattering events with nuclei~\cite{Andreopoulos:2009rq}. 
In this code, the boosted dark matter can also be implemented by slightly modifying the source code. 
We investigate the sensitivity at the DUNE experiment assuming 40 kton liquid argon detector with 10 years period of time~\cite{DUNE:2015lol}. 

The paper is organized as follows. 
In Section~\ref{sec:setup}, we describe the setup for our calculations, and estimate the theoretically expected number of signal events and atmospheric neutrino backgrounds. 
In Section~\ref{sec:event}, we illustrate how to reconstruct the signal events, and indicate the kinematic cuts, the energy/momentum thresholds and resolutions of the DUNE detector. 
Section~\ref{sec:result} is devoted to present the numerical results and interpretations. 
We summarize our work in Section~\ref{sec:summary}.

\section{Setup}
\label{sec:setup}

We consider the dark matter semi-annihilation $\chi\chi\to\nu\overline{\chi}$ where the dark matter $\chi$ must be a Dirac fermion 
to allow this process based on the spin conservation~\cite{Toma:2021vlw}. 
In general, the CP conjugate process $\overline{\chi}\overline{\chi}\to\overline{\nu}\chi$ is also allowed. 
However we concentrate only on $\chi\chi\to\nu\overline{\chi}$ to avoid unnecessary complication (namely the dark matter is asymmetric) though it is not a requirement in the following discussion. 
In fact, a possibility of generating such asymmetry via semi-annihilations has been proposed in a concrete model~\cite{Ghosh:2020lma}.
Hereafter the anti-dark matter in the final state is simply regarded as dark matter 
because this simplification does not lead any substantial difference.\footnote{Difference may arise for neutrinos because the interactions for neutrinos and anti-neutrinos are different.} 
Moreover, we assume that the dark matter abundance is thermally fixed by the semi-annihilation, and forms the dark matter halo. 
Namely the magnitude of the cross section is $\langle\sigma_\mathrm{semi}{v}_\mathrm{rel}\rangle\sim10^{-26}~\mathrm{cm^3/s}$. 

The dark matter particles are accumulated at the centre of the Sun due to the gravitational force. 
Then, the semi-annihilation $\chi\chi\to \nu\overline{\chi}$ and the capture process with nucleons $N$: $\chi N\to\chi N$ easily equilibrate 
if the dark matter mass is $m_\chi\gtrsim4~\mathrm{GeV}$~\cite{Garani:2017jcj}. 
Therefore we can anticipate the simultaneous signals of the high energy neutrino and boosted dark matter from the Sun. 
Because the dark matter particles accumulated in the Sun are non-relativistic, the energies of the neutrino and boosted dark matter in the final state 
are kinematically fixed to be $E_\nu=3m_\chi/4$ and 
$E_{\chi}=5m_\chi/4$, and the corresponding speed of the boosted dark matter is $v_\chi=0.6$. 
We will explore these two kinds of signals at the DUNE experiment with $40$~kton liquid argon fiducial volume and 10 years exposure~\cite{DUNE:2020ypp}. 
The DUNE experiment may not be able to detect the particles with the energy larger than $\mathcal{O}(100)~\mathrm{GeV}$ due to the detector design~\cite{DUNE:2015lol}. 
Therefore we concentrate on the dark matter mass range: $4~\mathrm{GeV}\leq m_\chi \leq 100~\mathrm{GeV}$ in this work. 

The concrete ultra-violet complete models leading the semi-annihilation process $\chi\chi\to\nu\overline{\chi}$ have been studied in refs.~\cite{Ma:2007gq, Aoki:2014cja} for example. 
In addition, similar discussions can be done for another semi-annihilation process $\chi\chi\to J\overline{\chi}$ in a concrete model~\cite{Miyagi:2022gvy} 
where $J$ is a Majoron being the Nambu-Goldstone boson associated with the global lepton number symmetry. The produced Majoron subsequently decays into neutrinos $J\to \nu\nu$. 
Model building is out of scope in this work, and we concentrate on the calculation of the signal events with the parametrization of the cross section with nucleons ($\chi N\to \chi N$)
as will be given in Eq.~(\ref{eq:xsec}) later though we will briefly comment how to construct the models.

\subsection{Neutrino signal and background}

\begin{figure}[t]
\begin{center}
\includegraphics[width=7.5cm]{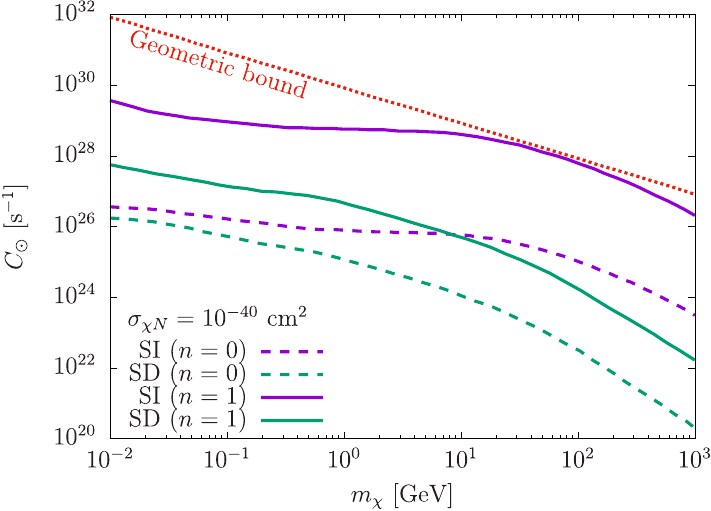}
\caption{Dark matter capture rate in the Sun with $\sigma_{\chi N}=10^{-40}~\mathrm{cm}^2$~\cite{Garani:2017jcj}. 
The solid purple and green lines represent the capture rates for spin-independent (SI) and spin-dependent (SD) cross sections with $Q^2$ dependence ($n=1$), respectively.  
The dashed purple and green lines are for constant cross sections ($n=0$).
The index $n=0,1$ labels the dependence of the momentum transfer $Q^2$ (See Eq.~(\ref{eq:xsec})).
The red dotted line in the top is the maximum capture rate, which is set by the geometrical cross section of the Sun.}
\label{fig:capture_rate}
\end{center}
\end{figure}

The neutrino flux produced by the dark matter semi-annihilation $\chi\chi\to\nu\overline{\chi}$ coming from the Sun, is given by 
\begin{align}
\frac{d^2\Phi_{\nu}}{dE_{\nu}d\Omega}=\frac{\Gamma_\mathrm{ann}}{4\pi d_\odot^2}\delta\left(E_{\nu}-\frac{3}{4}m_\chi\right)\delta\left(\Omega-\Omega_\odot\right),
\label{eq:1}
\end{align}
where $d_\odot=1.50\times10^{13}~\mathrm{cm}$ is the distance between Earth and the Sun, $\Omega_\odot$ is the Sun's solid angle, and $\Gamma_\mathrm{ann}$ is the semi-annihilation rate 
which is simply related with the capture rate in the Sun $C_\odot$ as $\Gamma_\mathrm{ann}=C_\odot/2$ 
under the assumption that the dark matter capture and semi-annihilation processes equilibrate in the Sun~\cite{Baratella:2013fya}. 
We consider the case that three flavors of neutrinos are equally produced at the Sun for simplicity. 
In this case, note that neutrino oscillations do not change the flavor ratio at Earth.\footnote{If the neutrino fluxes produced at the Sun are different for each flavor, neutrino oscillations change the fluxes at Earth and must be taken into account~\cite{Chauhan:2023zuf}.}
For the case that the $\chi N \to \chi N$ scattering cross section is a constant or is proportional to the momentum transfer $Q^2$, 
the capture rates have been calculated in ref.~\cite{Garani:2017jcj} as shown in Fig.~\ref{fig:capture_rate} 
where $\sigma_{\chi N}=10^{-40}~\mathrm{cm}^2$ is assumed for the non-relativistic dark matter velocity $v_\chi\sim10^{-3}$, 
and the capture rate simply scales as $C_\odot\propto \sigma_{\chi N}$. 
One can easily understand that the capture rate for the $Q^2$ dependent case is larger than the constant case because the dark matter velocity is accelerated by the Sun's gravitational force 
when the dark matter particles are captured. 
We also consider the $Q^4$ dependent case 
whose rate is simply obtained by scaling the $Q^2$ dependent one with the factor $590$ for the SI, and $50$ for SD cross sections. 
These factors are obtained by a simple extrapolation based on the same scaling between the constant and the $Q^2$ dependent rates as can be seen in Fig.~\ref{fig:capture_rate}, which has been discussed in ref.~\cite{Busoni:2017mhe}. 
Note that the rate cannot exceed the upper bound indicated by the red dotted line (Geometric bound) in Fig.~\ref{fig:capture_rate}~\cite{Garani:2017jcj}. 
This bound is simply determined by the geometry of the Sun and is independent of the elastic scattering cross section $\sigma_{\chi N}$.

\begin{figure}[t]
\begin{center}
\includegraphics[width=7.5cm]{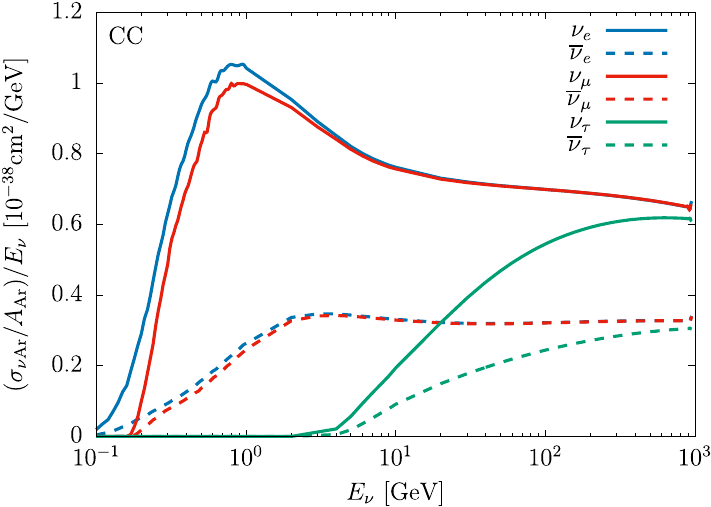}
\includegraphics[width=7.5cm]{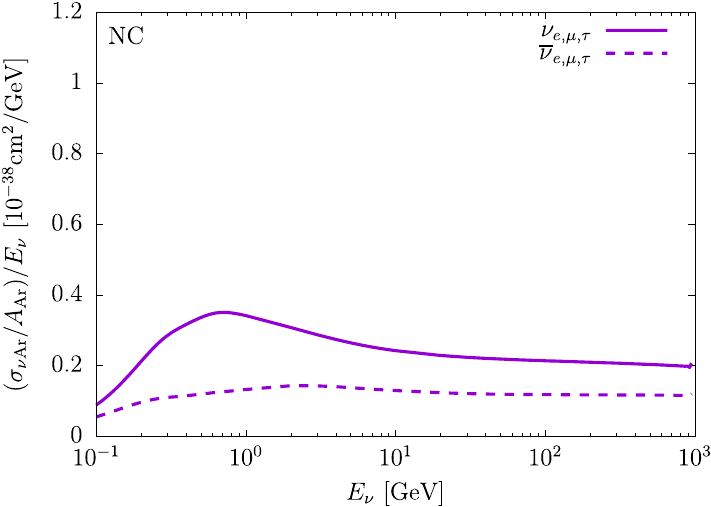}
 \caption{Neutrino-nucleon scattering cross sections in argon for CC (left) and NC (right) interactions~\cite{Andreopoulos:2015wxa} where $A_\mathrm{Ar}=39.9$ is the argon mass number.}
\label{fig:nu_xsec}
\end{center} 
\end{figure}

The high energy neutrino produced by the semi-annihilation $\chi\chi\to \nu\overline{\chi}$ enters in the DUNE detector and scatters off the nucleons in argon nuclei 
via the charged-current (CC) interaction. 
As a result, a charged lepton and some number of nucleons (a jet) are generated due to hadronization. 
Thus a pair of a charged lepton and a jet can be regarded as the signal event. 
The number of neutrino signal events is calculated as 
\begin{align}
N_{\nu}^\mathrm{CC}=N_NT\sum_{\alpha}\int \sigma_{\nu_\alpha N}^\mathrm{CC}\frac{d^2\Phi_{\nu_\alpha}}{dE_{\nu_\alpha} d\Omega}dE_{\nu_\alpha}d\Omega,
\label{eq:num_nu}
\end{align}
where $\alpha$ is the neutrino flavor index ($e,\overline{e},\mu,\overline{\mu}$), 
$N_N=2.41\times10^{34}$ is the number of nucleons in liquid argon target (assuming $40$ kton fiducial volume), $T=10~\mathrm{yr}$ is the period of the experiment, 
$\sigma_{\nu_\alpha N}^\mathrm{CC}$ is the neutrino-nucleon scattering cross section for argon via the CC interaction, and 
$d^2\Phi_{\nu_\alpha}/dE_{\nu_\alpha} d\Omega$ is the differential neutrino flux for a flavor $\alpha$, which is given by $1/3$ of Eq.~(\ref{eq:1}) because three flavors of neutrinos are equally produced. 
Integrating over the energy and the solid angle, we obtain
\begin{align}
N_{\nu}^\mathrm{CC}=\frac{N_NT}{3}\left.\sum_{\alpha}\frac{C_\odot}{8\pi d_\odot^2}\sigma_{\nu_\alpha N}^\mathrm{CC}\right|_{E_{\nu_\alpha}=3m_\chi/4}.
\label{eq:nu_signal}
\end{align}
Precise evaluation of the cross section $\sigma_{\nu_\alpha N}^\mathrm{CC}$ is difficult especially around $E_{\nu}=\mathcal{O}(1)~\mathrm{GeV}$ because of the complicated nuclear interactions. 
Here we adopt the evaluation based on the experimental nuclear model implemented in \texttt{GENIE}~\cite{Andreopoulos:2009rq}. 
The extracted CC (and neutral-current (NC)) cross sections in argon target are shown in Fig.~\ref{fig:nu_xsec}. 
One can easily confirm that these CC cross sections are consistent with the various experimental measurements~\cite{ParticleDataGroup:2018ovx}. 

The main background for the neutrino signal is atmospheric neutrinos ($\nu_\mathrm{atm} + N \to \ell + j$) whose number of events for each flavor is calculated as
\begin{align}
N_{\mathrm{atm}\hspace{0.05cm}\nu_\alpha}^\mathrm{CC}=N_NT\int \sigma_{\nu_{\alpha} N}^{\mathrm{CC}}\frac{d^2\Phi_{\nu_\alpha}^\mathrm{atm}}{dE_{\nu_\alpha} d\Omega}dE_{\nu_\alpha}d\Omega,
\end{align}
where 
$d^2\Phi_{\nu_\alpha}^{\mathrm{atm}}/dE_{\nu_\alpha}d\Omega$ is the atmospheric neutrino flux for the neutrino flavor $\alpha$. 
We use the atmospheric neutrino fluxes at Homestake based on the HAKKM2014 model~\cite{Honda:2015fha, Super-Kamiokande:2015qek}, 
and average the fluxes with the minimum and maximum solar modulation effect. 
The effect of neutrino oscillations are included in the HAKKM2014 model. 
We are interested in the flux coming from the Sun's direction, and thus we estimate the effective solid angle as $\Delta\Omega = \pi\tan(2^\circ+10^\circ) = 0.668$ taking into account double of the DUNE angular resolution ($1^\circ$ for charged leptons and $5^\circ$ for the others as will be seen in Tab.~\ref{tab:1}).\footnote{The angular resolutions at DUNE could be worse in more sophisticated analysis~\cite{DUNE:2021gbm}. We expect a factor difference in our final results if more conservative angular resolutions are assumed.}
We have done this treatment because the DUNE detector angular resolution is much larger than the actual Sun's solid angle.
Thus the number of events for the atmospheric neutrino background coming from the Sun is evaluated as
\begin{align}
N_{\mathrm{atm}\hspace{0.05cm}\nu_\alpha}^\mathrm{CC}=N_NT\frac{\Delta\Omega}{4\pi}\int \sigma_{\nu_{\alpha} N}^{\mathrm{CC}}\frac{d\Phi_{\nu_\alpha}^\mathrm{atm}}{dE_{\nu_\alpha}}dE_{\nu_{\alpha}}.
\label{eq:num_nuatm}
\end{align}
Integrating over the atmospheric neutrino energy, 
the expected number of the atmospheric neutrino background for each flavor is numerically estimated as
\begin{align}
N_{\mathrm{atm}\hspace{0.05cm}\nu_e}^\mathrm{CC}=612.1,\quad 
N_{\mathrm{atm}\hspace{0.05cm}\nu_\mu}^\mathrm{CC}=119.5,\quad
N_{\mathrm{atm}\hspace{0.05cm}\overline{\nu}_e}^\mathrm{CC}=1077,\quad 
N_{\mathrm{atm}\hspace{0.05cm}\overline{\nu}_\mu}^\mathrm{CC}=260.6,
\end{align}
for 40 kton liquid argon and 10 years exposure, and the total number of events is $2070$. 
We take into account only electron and muon flavors of neutrinos, and tau flavor is not considered here. 
Regarding tau neutrino background, there is an argument in Ref.~\cite{Berger:2019ttc}.

\subsection{Boosted dark matter signal and background}
For the boosted dark matter signal, there are three kinds of relevant processes which are (quasi-)elastic, 
resonant and deep inelastic scattering (DIS) processes~\cite{Berger:2018urf, Berger:2019ttc}. 
The dominant process changes depending on the magnitude of the momentum transfer $Q^2$ which is related with the Mandelstam variable as $t=-Q^2$. 
For the semi-annihilation process $\chi\chi\to \nu \overline{\chi}$ we focus on, the dark matter velocity is fixed at $v_\chi=0.6$ by kinematics. 
In this case, the range of the momentum transfer is approximately limited in $Q^2 \lesssim 9m_N^2/4\approx \left(1.4~\mathrm{GeV}\right)^2$ for the relevant processes ($m_N\ll m_\chi$). 
In this energy range, the (quasi-)elastic scattering is dominant over the other processes. 

We parametrize the differential elastic scattering cross section for $\chi N\to\chi N$ as~\cite{Garani:2017jcj}
\begin{align}
\frac{d\sigma_{\chi N}}{dQ^2}&=\frac{\sigma_0s}{4m_N^2|\bm{p}_\chi|^2}\left(\frac{Q^2}{m_N^2v_0^2}\right)^n|F(Q^2)|^2,
\label{eq:xsec}
\end{align}
where $s$ is the Mandelstam variable, 
$|\bm{p}_\chi|=3m_\chi/4$ is the initial dark matter momentum, 
$v_0=220~\mathrm{km/s}$ is a reference speed of dark matter, 
and $\sigma_0$ is a reference cross section. 
The index $n$ represents the order of momentum transfer dependence and we take $n=0,1,2$. 
For the form factor $F(Q^2)$, we adopt a dipole form factor described by~\cite{Berger:2019ttc}
\begin{align}
F(Q^2)=\frac{F(0)}{\left(1+Q^2/(0.99~\mathrm{GeV})^2\right)^2},
\end{align}
where the normalization factor can be regarded as $F(0)=1$ because it is absorbed by the reference cross section $\sigma_0$.\footnote{In fact, the shape of the form factor depends on unknown dark matter interactions.}  
The total scattering cross section can be obtained by integrating in the range $0<Q^2<\lambda(s,m_\chi^2,m_N^2)/s$ where 
$\lambda(x,y,z)=x^2+y^2+z^2-2xy-2yz-2xz$ is the kinematic function (K\"{a}ll\'{e}n function). 
Note that the differential cross section in Eq.~(\ref{eq:xsec}) is related to the non-relativistic elastic cross section $\sigma_{\chi N}^0$ relevant to dark matter direct detection as 
\begin{align}
\sigma_{\chi N}^0=\frac{\sigma_0}{n+1}\left(\frac{2m_\chi}{m_\chi+m_N}\right)^{2n},
\end{align}
which is obtained by integrating over $Q^2$ with $F(Q^2)=1$ and $|\bm{p}_\chi|=m_\chi v_0$.

The non-trivial momentum transfer dependence on the differential cross section in Eq.~(\ref{eq:xsec}) can be implemented as follows~\cite{Gelmini:2018ogy}. 
One can build $Q^2$ dependent models by considering the scalar--pseudo-scalar, pseudo-scalar--scalar or anapole moment interactions: 
\begin{align}
\mathcal{L}_{SP} & = -y_{\chi}^{S} \varphi \overline{\chi}\chi - y_{q}^{P} \varphi \overline{q} \gamma_5 q,\\
\mathcal{L}_{PS} & = -y_{\chi}^{P} \varphi \overline{\chi}\gamma_5 \chi - y_{q}^{S}\varphi \overline{q} q,\\
\mathcal{L}_\mathrm{ana} & = a_\chi \overline{\chi}\gamma_{\mu}\gamma_5 \partial_{\nu}\chi F^{\mu\nu} - eA_{\mu}\overline{q}\gamma^{\mu}q,
\end{align}
where $\varphi$ is the mediator between the dark matter and nucleons (quarks) scattering, $F^{\mu\nu}$ is the electromagnetic field strength, and $A_{\mu}$ is the electromagnetic field. 
The first model gives a SD cross section while the second model is a SI cross section. 
The $Q^4$ dependent models can be built with the pseudo-scalar--pseudo-scalar interaction:
\begin{align}
\mathcal{L}_{PP} & = - y_{\chi}^{P} \varphi \overline{\chi}\gamma_5\chi - y_{q}^{P} \varphi \overline{q} \gamma_5 q.
\end{align}
Note that one may need to take into account the constraint from indirect detection via $\chi\overline{\chi}\to q\overline{q}$ if the dark matter abundance in our galaxy is not asymmetric completely. 

Similar to the neutrino signal in the previous subsection, the number of the boosted dark matter signal events is calculated as 
\begin{align}
N_{\chi}=N_NT\int \sigma_{\chi N}\frac{d^2\Phi_{\chi}}{dE_\chi d\Omega}dE_{\chi}d\Omega,
\end{align}
where $d^2\Phi_\chi/dE_{\chi}d\Omega$ is the boosted dark matter flux produced by the dark matter semi-annihilation $\chi\chi\to\nu\overline{\chi}$, which can be given by 
\begin{align}
\frac{d^2\Phi_{\chi}}{dE_{\chi}d\Omega}=\frac{\Gamma_\mathrm{ann}}{4\pi d_\odot^2}\delta\left(E_{\chi}-\frac{5}{4}m_\chi\right)\delta\left(\Omega-\Omega_\odot\right).
\end{align}
Integrating over the energy and the solid angle, we obtain
\begin{align}
N_{\chi}=N_NT\left.\frac{C_\odot}{8\pi d_\odot^2}\sigma_{\chi N}\right|_{E_\chi=5m_\chi/4}.
\label{eq:chi_signal}
\end{align}

The main background for the boosted dark matter signal is atmospheric neutrinos via the NC interaction ($\nu_\mathrm{atm} + N \to \nu_\mathrm{atm} + N$). 
The number of events for the background is calculated as
\begin{align}
N_{\mathrm{atm}\hspace{0.05cm}\nu_\alpha}^\mathrm{NC}=N_NT\int \sigma_{\nu_{\alpha} N}^{\mathrm{NC}}\frac{d^2\Phi_{\nu_\alpha}^\mathrm{atm}}{dE_{\nu_\alpha} d\Omega}dE_{\nu_\alpha}d\Omega,
\label{eq:num_chi_atm}
\end{align}
where $\sigma_{\nu_\alpha N}^\mathrm{NC}$ is the neutrino--nucleon cross section in argon nuclei via the NC interaction, 
which is extracted from \texttt{GENIE} and shown in the right panel of Fig.~\ref{fig:nu_xsec}. 
Integrating over the atmospheric neutrino energy, the expected number of the background for each flavor 
is estimated as
\begin{align}
N_{\mathrm{atm}\hspace{0.05cm}\nu_e}^\mathrm{NC}=240.5,\quad 
N_{\mathrm{atm}\hspace{0.05cm}\nu_\mu}^\mathrm{NC}=82.65,\quad
N_{\mathrm{atm}\hspace{0.05cm}\overline{\nu}_e}^\mathrm{NC}=477.9,\quad 
N_{\mathrm{atm}\hspace{0.05cm}\overline{\nu}_\mu}^\mathrm{NC}=193.4,
\end{align}
for 40 kton liquid argon and 10 years exposure, and the total number of events is $994.3$. 
The monochromatic neutrino signals with $E_{\nu}=3m_\chi/4$ discussed in the previous subsection may also be a background for the boosted dark matter signal. 
However, the dominant process is the DIS in this case because the neutrino energy is much larger than $\mathcal{O}(1)~\mathrm{GeV}$, and we can easily discard these events from the background. 
The other background events could arise from misidentification of CC neutrino scatterings for example. 
We interpret that these effects are taken into account as the systematic uncertainty in the calculation of DUNE sensitivities as we will see later in Eqs.~(\ref{eq:sig1}) and (\ref{eq:sig2}). 

From the above argument, one can see that the number of events for the boosted dark matter is roughly proportional to the square of the elastic scattering cross section 
between dark matter and nucleons as in Eq.~(\ref{eq:chi_signal}) (the capture rate is also proportional to $\sigma_{\chi N}$). 
On the other hand, the number of events for the neutrino signal is simply proportional to the elastic scattering cross section coming from the capture rate as can be seen Eq.~(\ref{eq:nu_signal}).

\section{Event reconstruction}

\label{sec:event}
We generate the neutrino and boosted dark matter signals using \texttt{GENIE}~\cite{Andreopoulos:2009rq}, and investigate if those signals are detected at the DUNE detector. 
The detector threshold, energy/momentum resolution and angular resolution are shown in Tab.~\ref{tab:1}.\footnote{The angular resolution may depend on the energy and channels of interest such as CC electron or CC muon neutrino events. If such dependence is taken into account in more sophisticated analysis~\cite{DUNE:2021gbm}, it could give a considerable effect on the detection sensitivity.}
We apply the detector information in Tab.~\ref{tab:1} for the event reconstructions of the NC neutrino scattering and boosted dark matter scattering.
For the energy resolution of CC neutrino events, we adopt the value based on the detailed event reconstruction~\cite{Friedland:2018vry}. The energy dependence of neutrino energy resolution has been estimated in the reference up to $5~\mathrm{GeV}$. We extrapolate the energy dependence up to $100~\mathrm{GeV}$ with a simple power low for our purpose.
The events below the threshold will be discarded, and the generated observables such as a momentum/energy and angle will be smeared with the energy/momentum and angular resolutions. 
Then we regard the event is observed only if the reconstructed energy after smearing is within the 2$\sigma$ energy resolution. 
In the following, we show how to reconstruct the signals from the observed quantities. 

Furthermore, if one would like to take into account the seasonal modulation of the signal events from the Sun, 
the public code \texttt{GenSolFlux}~\cite{Berger:2018urf} and \texttt{SolTrack}~\cite{soltrack} can be used though we do not consider in our work.

\begin{table}[t]
\begin{center}
\begin{tabular}{|c||c|c|c|}\hline
 & Detector threshold & Energy/momentum resolution & Angular resolution\\
\hhline{|=#=|=|=|}
$\mu^\pm$ & 30 MeV & $5$ \% & $1^\circ$ \\\hline
$\pi^\pm$ & 100 MeV & $5$ \% & $1^\circ$ \\\hline
$e^\pm/\gamma$ & 30 MeV & $2+15/\sqrt{E/\mathrm{GeV}}$ \% & $1^\circ$ \\\hline
\multirow{2}{*}{$p$} & \multirow{2}{*}{50 MeV} & $|\bm{p}|<400$ MeV: $10$ \%\hspace{2.7cm} & \multirow{2}{*}{$5^\circ$} \\
                     &                         & $|\bm{p}|>400$ MeV: $5+30/\sqrt{E/\mathrm{GeV}}$ \% & \\\hline
$n$ & 50 MeV & $40/\sqrt{E/\mathrm{GeV}}$ \% & $5^\circ$ \\\hline
\end{tabular} 
\caption{Detector threshold, energy/momentum resolution and angular resolution of the DUNE detector~\cite{DUNE:2015lol}.}
\label{tab:1}
\end{center}
\end{table}

\subsection{Neutrino energy reconstruction}
\begin{figure}[t]
\begin{center}
\includegraphics[width=6.5cm]{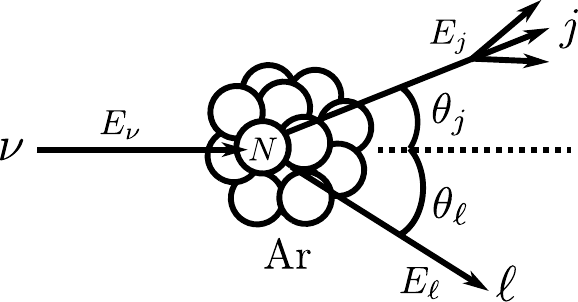}
\hspace{1cm}
\includegraphics[width=6.5cm]{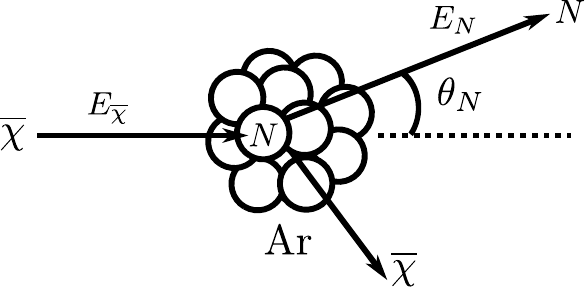}
\caption{(Left): momenta and angles of the charged lepton $\ell$ and jet $j$ in the final state. These quantities can be identified with good precision at DUNE.
(Right): kinematics for the (quasi-)elastic scattering of the boosted dark matter signal.}
\label{fig:kinematics}
\end{center}
\end{figure}

The high energy neutrinos produced by the dark matter semi-annihilation generate a charged lepton and a jet via the CC interaction at the DUNE detector 
as illustrated in the left panel of Fig.~\ref{fig:kinematics}
where the jet is constructed by several number of nucleons, charged pions and gamma. 
The energy of the charged lepton ($E_\ell$) and the direction of the charged lepton ($\theta_\ell$) and jet ($\theta_j$) 
can be identified with good precision at DUNE while the jet energy is not so precise. 
Using these observed quantities, one can reconstruct the neutrino energy as~\cite{Rott:2019stu} 
\begin{align}
E_{\nu}=\frac{1}{2}\frac{\sin\theta_j\left(1+\cos\theta_\ell\right) + \sin\theta_\ell\left(1+\cos\theta_j\right)}{\sin\theta_j}E_{\ell}.
\end{align}
The angles $\theta_\ell$ and $\theta_j$ are expected to be very small for the neutrino energy $E_{\nu}\gtrsim\mathcal{O}(1)~\mathrm{GeV}$. 
Then, the reconstructed neutrino energy via the CC interaction for each event is smeared with the energy resolution~\cite{Friedland:2018vry}.

For the NC neutrino events which are the main background for the boosted dark matter signal, 
the neutrino energy reconstruction is done as same with the boosted dark matter energy reconstruction as will be explained in the next subsection where we adopt the energy and angular resolutions in Tab.~\ref{tab:1} for the final state particles.

\subsection{Boosted dark matter energy reconstruction}
For the boosted dark matter, the scattering kinematics is simply determined because the main process is (quasi-)elastic scattering $\chi N\to \chi N$ in our case. 
We define the recoil nucleon energy $E_N$ and the angle $\theta_N$ measured from the solar direction as shown in the right panel of Fig.~\ref{fig:kinematics}. 
Thus using the energy/momentum conservation, the boosted dark matter energy $E_\chi$ can be reconstructed for the elastic scattering as
\begin{equation}
E_{\chi}=m_N\frac{1+\alpha\cos\theta_N\sqrt{1-\beta + \alpha^2\beta\cos^2\theta_N}}{-1+\alpha^2\cos^2\theta_N },
\end{equation}
where $\alpha=\sqrt{(E_N+m_N)/(E_N-m_N)}>1$ and $\beta=m_\chi^2/m_N^2>1$.
Therefore the incoming boosted dark matter energy can be reconstructed from the observables $E_N$ and $\theta_N$. 
Because the velocity of the boosted dark matter is moderate ($v_\chi=0.6$), the scattering angle $\theta_N$ is widely spread unlike the neutrino case in the above. 
The reconstructed boosted dark matter energy with smearing of the nucleons in the final state as in Tab.~\ref{tab:1} is compared to the true energy. 
Then we judge if the reconstructed energy is in $2\sigma$ range of the dark matter energy resolution which is regarded as the same with neutrinos~\cite{Friedland:2018vry}.

For the quasi-elastic scattering, several number of nucleons may be produced in the final state due to hadronization. 
We take the events with only one nucleon in the final state because the resolution uncertainty becomes larger for the case that several number of nucleons are produced. 
We have numerically checked that approximately $32$\% of the quasi-elastic scattering events involve only one nucleon. 
We expect such discrimination can easily be done because the reconstructed boosted dark matter energy for the multi-nucleons case is completely different from the true energy which is theoretically determined.
For comparison, we have also checked what happen if the number of nucleons in the final state is not restricted as above. 
It has been turned out that the signal significance which will be defined by Eq.~(\ref{eq:sig2}) becomes worse $30~\%$ or less than the single nucleon production case.

\section{Results}
\label{sec:result}

\subsection{Benchmark parameter sets}
\begin{table}[t]
\begin{center}
\begin{tabular}{|c||c|c|c|c|c|}\hline
 & model & $m_\chi$ [GeV] & $\sigma_0~[\mathrm{cm}^2]$ & \# of $\nu$ events & \# of $\chi$ events \\\hhline{|=#=|=|=|=|=|}
\multirow{2}{*}{BP1} & \multirow{2}{*}{SD ($n=1$)} & \multirow{2}{*}{$6$} & \multirow{2}{*}{$1.2\times10^{-42}$} & $N_{\mathrm{atm}\hspace{0.05cm}\nu}^\mathrm{CC}=54/2070$ & $N_{\mathrm{atm}\hspace{0.05cm}\nu}^\mathrm{NC}=98/994$ \\
                     &            &      &                                                                        & $N_{\nu}^\mathrm{CC}=18/47$ & $N_{\chi}=113/372$ \\\hline
\multirow{2}{*}{BP2} & \multirow{2}{*}{SD ($n=2$)} & \multirow{2}{*}{$30$} & \multirow{2}{*}{$5.0\times10^{-46}$} & $N_{\mathrm{atm}\hspace{0.05cm}\nu}^\mathrm{CC}=1/2070$ & $N_{\mathrm{atm}\hspace{0.05cm}\nu}^\mathrm{NC}=18/994$ \\
                     &            &      &                                                                        & $N_{\nu}^\mathrm{CC}=0/0$ & $N_{\chi}=405/2117$ \\\hline
\end{tabular}
\end{center}
\caption{Benchmark parameter sets and the observed/expected number of events after the cut with the detector threshold and resolution. For the number of events, the neutrino flavors ($\alpha=e,\overline{e},\mu,\overline\mu$) are summed.}
\label{tab:bp}
\end{table}

\begin{figure}[t]
\begin{center}
\includegraphics[width=7.5cm]{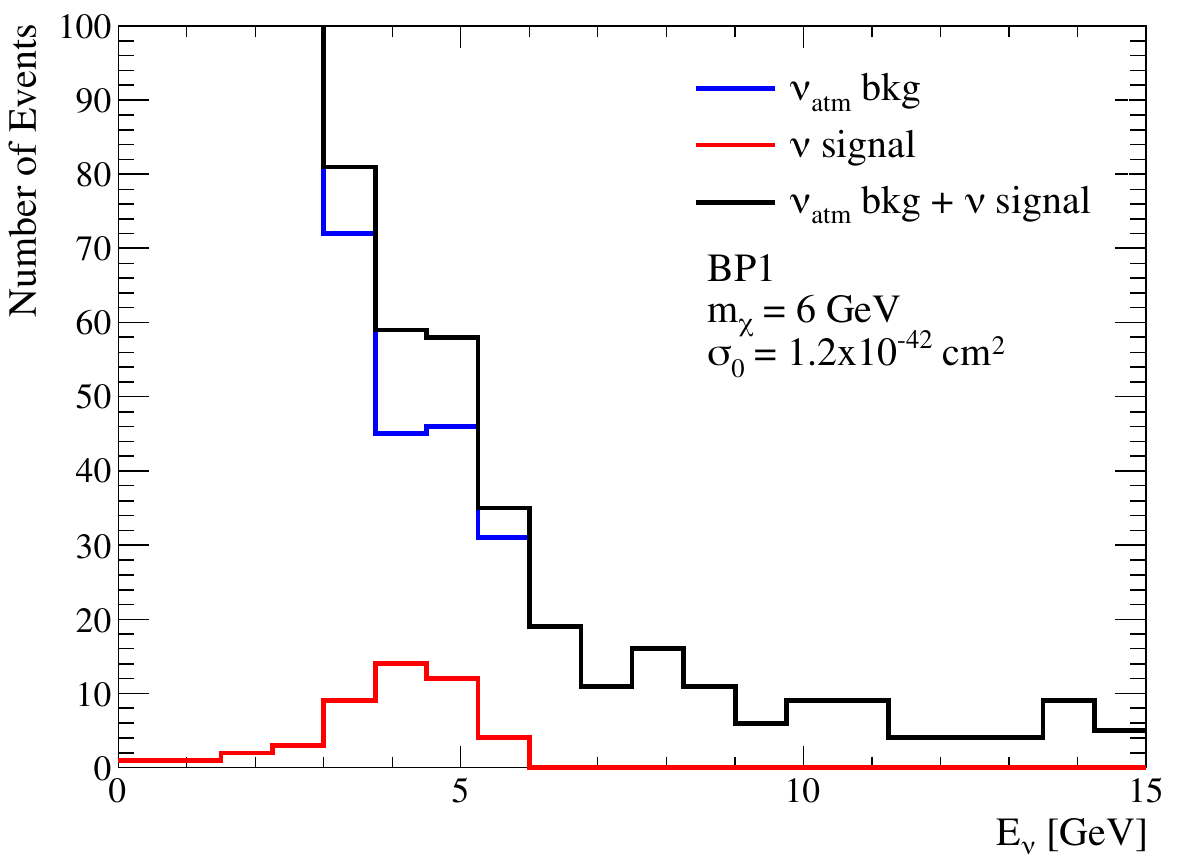}
\includegraphics[width=7.5cm]{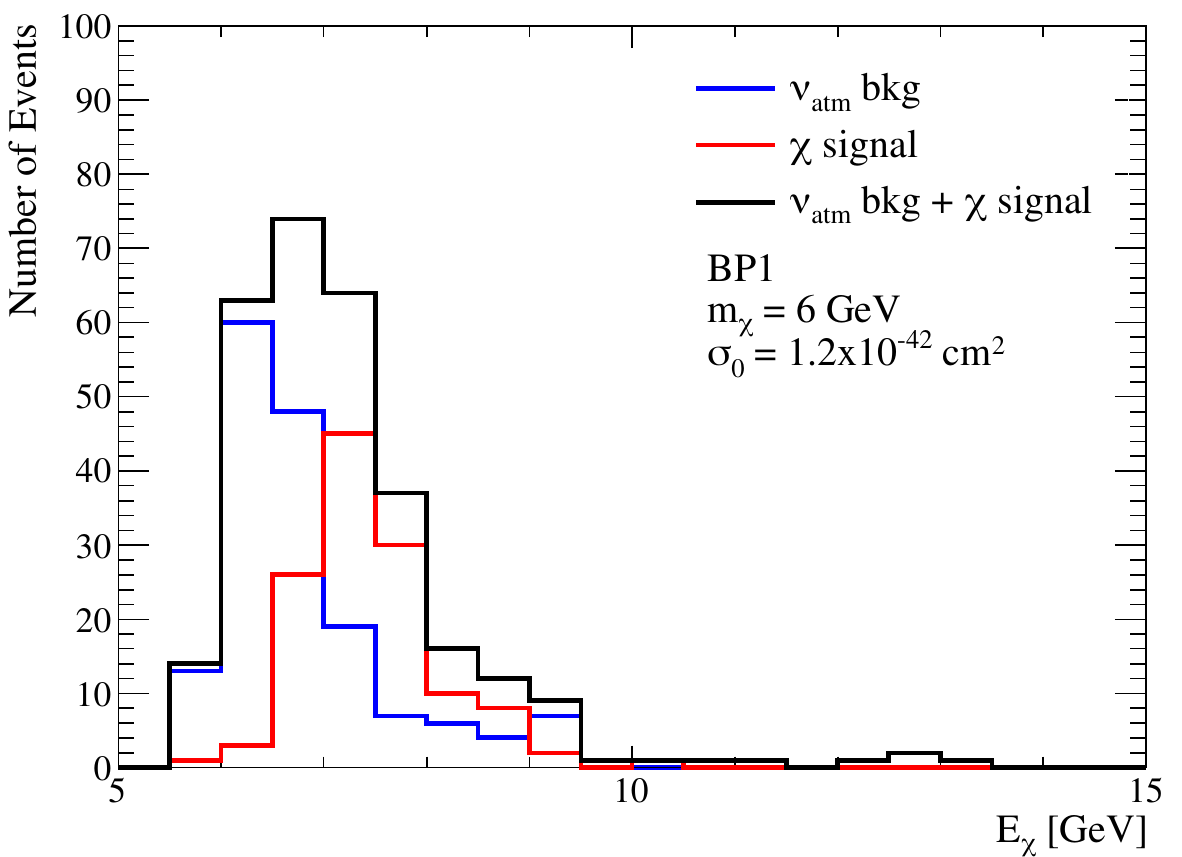}\\
\includegraphics[width=7.5cm]{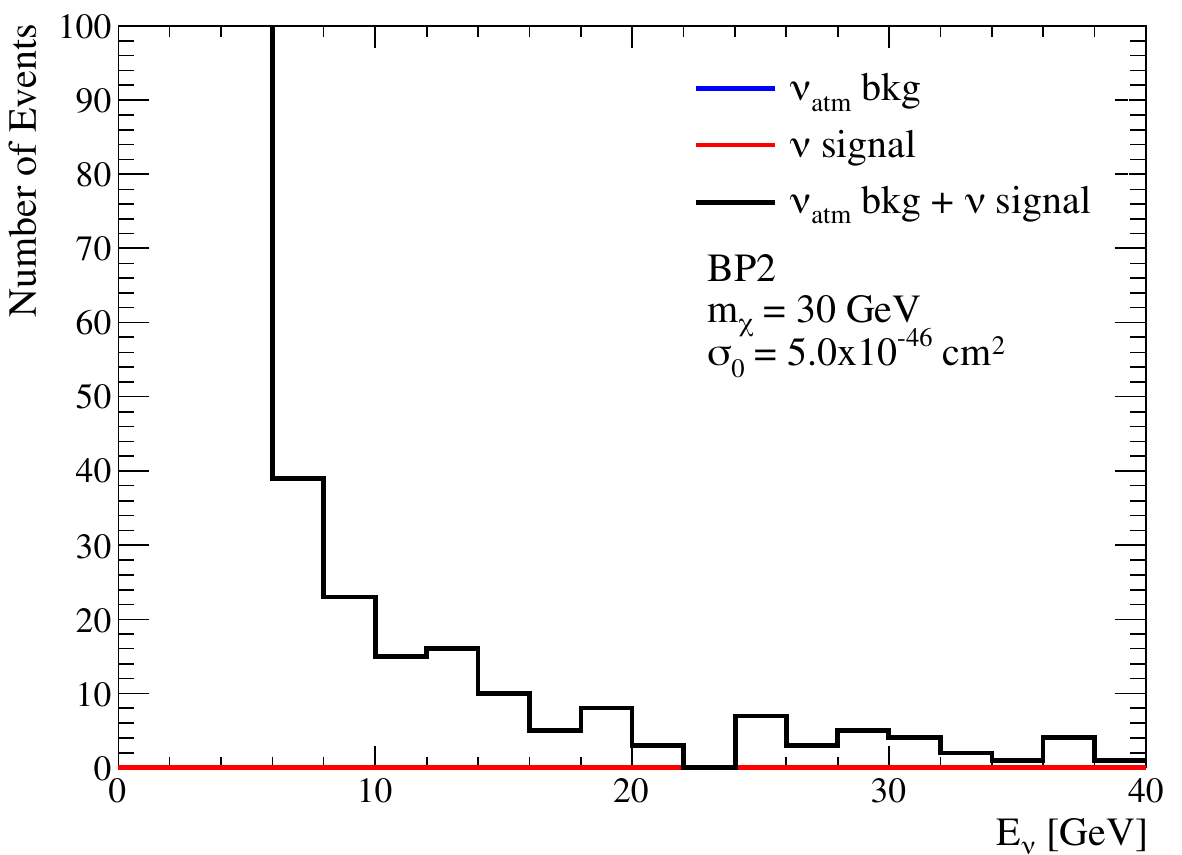}
\includegraphics[width=7.5cm]{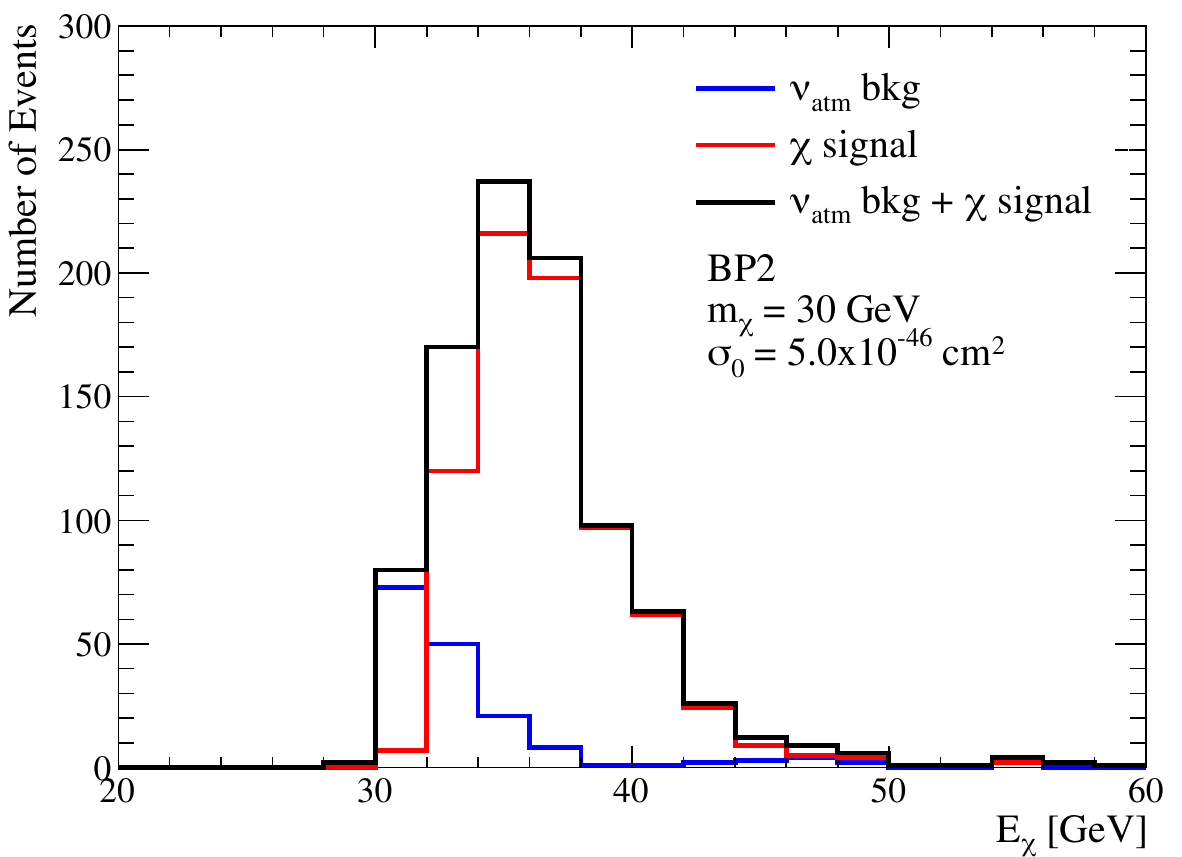}
\caption{Reconstructed neutrino energy distribution (left) and boosted dark matter energy distribution (right). 
The upper (lower) panels correspond to BP1 (BP2) in Tab.~\ref{tab:bp}.
 The blue, red and black lines represent the number of events for atmospheric neutrino background, neutrino/boosted dark matter signal and sum of them, respectively.}
\label{fig:dist}
\end{center} 
\end{figure}

Two benchmark parameter sets and the observed/expected number of events are shown in Tab.~\ref{tab:bp} where 
the expected numbers of events are the theoretical predictions calculated from Eqs.~(\ref{eq:num_nu}) and (\ref{eq:num_nuatm}) for the neutrinos and 
Eqs.~(\ref{eq:chi_signal}) and (\ref{eq:num_chi_atm}) for the boosted dark matter 
while the observed numbers of events are the rest of events after the discard with the detector thresholds and resolutions. 
The BP1 indicates relatively the small number of boosted dark matter signal events $N_\chi$ while the BP2 is a large number. 
This is because the $n=2$ model has a strong dependence of the momentum transfer, and the rate is highly enhanced. 
Fig.~\ref{fig:dist} shows the reconstructed energy distributions of the neutrino and boosted dark matter for BP1 (upper panels) and BP2 (lower panels). 
The blue, red and black lines correspond to the number of the atmospheric neutrino background, the neutrinos/boosted dark matter signals and sum of those events, respectively. 
The most of the reconstructed energy of the atmospheric neutrino background has a few GeV as expected from the energy flux. 
The true energies are $E_{\nu}=4.5~\mathrm{GeV}$ and $E_{\chi}=7.5~\mathrm{GeV}$ for the BP1 while those are $E_{\nu}=22.5~\mathrm{GeV}$ and $E_{\chi}=37.5~\mathrm{GeV}$ for the BP2. 
One can find from the plots that these true energies are reconstructed well with a dispersion. 
For BP2, since the observed/expected numbers of neutrino events are zero, it would be difficult to find the simultaneous signals of the semi-annihilation only at DUNE. 
However, it is possible that the boosted dark matter is found at DUNE while the neutrino signal is checked at the other neutrino experiments such as IceCube/DeepCore and Hyper-Kamiokande.

\subsection{Parameter space}
\begin{figure}[t]
\begin{center}
\includegraphics[width=7.5cm]{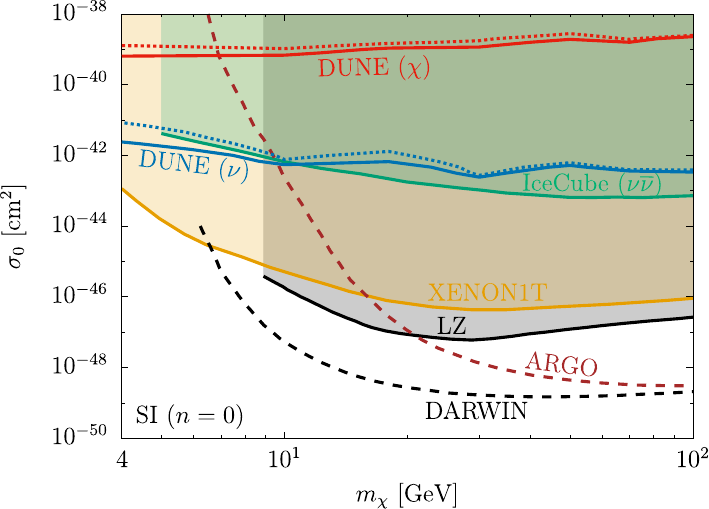}
\includegraphics[width=7.5cm]{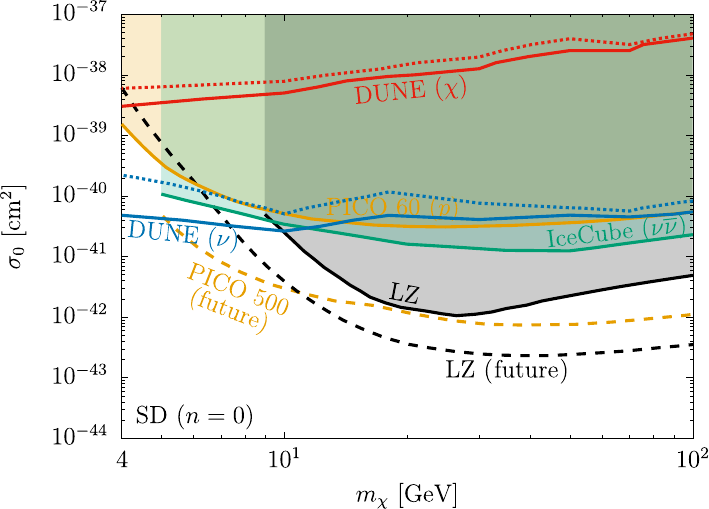}
\caption{
DUNE sensitivity at $2\sigma$ for the neutrino (blue) and boosted dark matter (red) signals ($n=0$). 
The existing upper bounds and future sensitivities for dark matter cross section $\sigma_0$ are also shown. 
The black and orange regions are excluded by the direct detection experiments (LZ~\cite{LZ:2022ufs}, XENON1T~\cite{XENON:2018voc} and PICO 60~\cite{PICO:2019vsc}) 
while the green region is excluded by the monochromatic neutrino observation at IceCube~\cite{IceCube:2021xzo}. 
The dashed black and brown lines on the left panel are the future sensitivities of DARWIN~\cite{Schumann:2015cpa, DARWIN:2016hyl} and ARGO~\cite{ARGO} 
while the dashed black and orange lines on the right panel are the future sensitivities of LZ~\cite{LZ:2018qzl} and PICO 500~\cite{Giroux:2021vpy}. 
}
\label{fig:parameter_space}
\end{center}
\end{figure}

We search for the parameter space testable by DUNE in ($m_\chi$, $\sigma_0$) plane. 
For this purpose, we define the signal significance
\begin{align}
\mathcal{S}_\nu &= \frac{N_\nu^\mathrm{CC}}{\sqrt{N_{\mathrm{atm}\hspace{0.05cm}\nu}^\mathrm{CC} + N_\nu^\mathrm{CC} + \delta_\nu^2}},\label{eq:sig1}\\
\mathcal{S}_\chi &= \frac{N_\chi}{\sqrt{N_{\mathrm{atm}\hspace{0.05cm}\nu}^\mathrm{NC} + N_\chi + \delta_\chi^2}},\label{eq:sig2}
\end{align}
for the neutrino and boosted dark matter signals, respectively. 
Here $\delta_\nu$ and $\delta_\chi$ are the systematic uncertainties given by
\begin{align}
\delta_\nu &=\left(N_{\mathrm{atm}\hspace{0.05cm}\nu}^\mathrm{CC} + N_\nu^\mathrm{CC}\right)\epsilon_\nu,\\
\delta_\chi &=\left(N_{\mathrm{atm}\hspace{0.05cm}\nu}^\mathrm{NC} + N_\chi\right)\epsilon_\chi.
\end{align}
The systematic uncertainties of atmospheric neutrinos at DUNE have been studied~\cite{Kelly:2019itm, Kelly:2021jfs, DeRomeri:2021xgy}. 
Following these references, we take the two values $\epsilon_{\nu}=\epsilon_{\chi}=0$ and $0.2$.\footnote{We checked that the signals can be explored at DUNE if $\epsilon_{\nu} = \epsilon_{\chi} \lesssim 0.4$.} 
For the constant cross section ($n=0$), the $2\sigma$ DUNE sensitivity is shown in Fig.~\ref{fig:parameter_space} for the SI (left) and SD (right) cross sections. 
The solid blue and red lines represent the DUNE sensitivities for the neutrino and boosted dark matter signals for no systematic uncertainties ($\epsilon_{\nu}=\epsilon_{\chi}=0$).
Namely, the above region from each line can be searched by DUNE. 
The dotted blue and red lines correspond to the sensitivities for $0.2$ systematic uncertainties. 
The black and orange regions are excluded by the direct detection experiments (LZ~\cite{LZ:2022ufs}, XENON1T~\cite{XENON:2018voc} and PICO 60~\cite{PICO:2019vsc}).
The dashed black (DARWIN~\cite{Schumann:2015cpa, DARWIN:2016hyl}) and brown (ARGO~\cite{ARGO}) lines on the left panel represent 
the future sensitivities of direct detection experiments for the SI case 
while the dashed black (LZ~\cite{LZ:2018qzl}) and orange (PICO 500~\cite{Giroux:2021vpy}) lines on the right panel correspond to the SD case.
The green region on the right panel is excluded by the monochromatic neutrino observation at IceCube~\cite{IceCube:2021xzo}. 
The green region for the left panel has been obtained by translating the original IceCube bound for the SD cross section into the SI cross section with the corresponding capture rates. 
Because the constraint of the direct detection experiments is very strong, all the parameter space for the boosted dark matter which can be tested by DUNE (the upper region from the red line) 
is completely excluded as expected. 

\begin{figure}[t]
\begin{center}
\includegraphics[width=7.5cm]{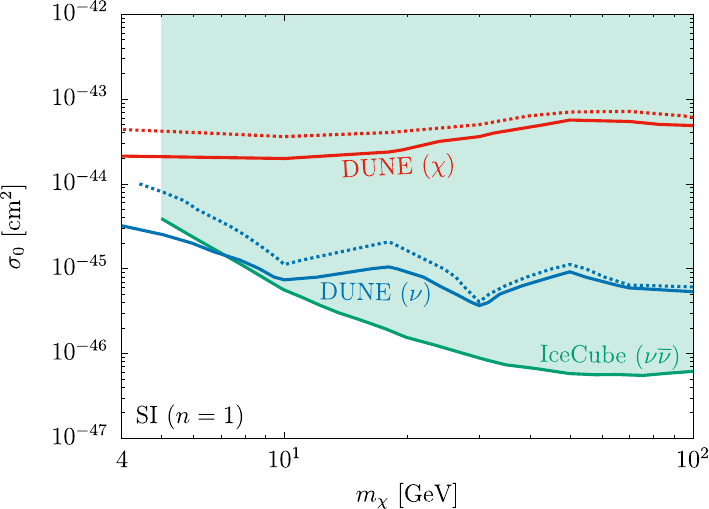}
\includegraphics[width=7.5cm]{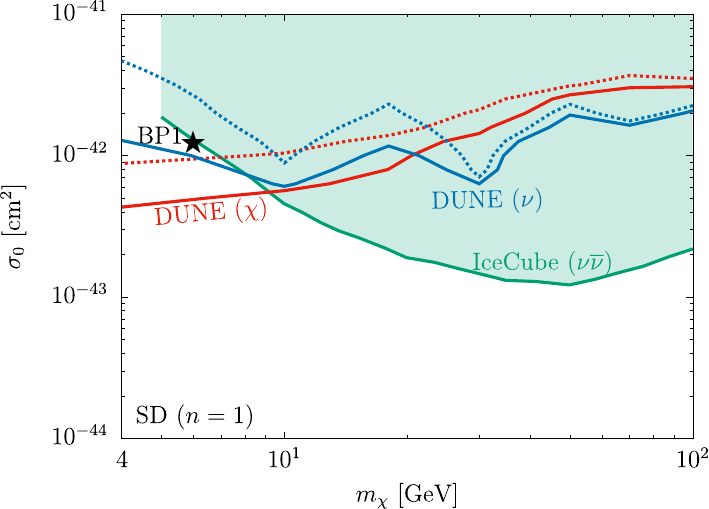}\\
\includegraphics[width=7.5cm]{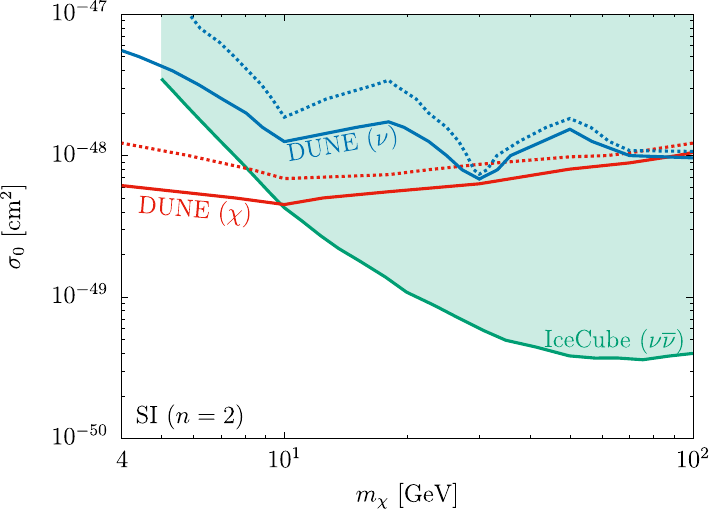}
\includegraphics[width=7.5cm]{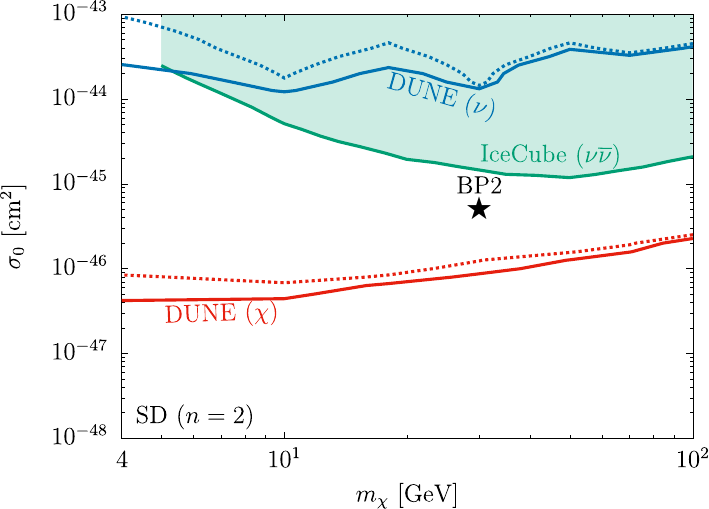}
\caption{
Same plots with Fig.~\ref{fig:parameter_space} for $n=1$ and $2$. The black stars represent the benchmark parameter sets (BP1 and BP2).}
\label{fig:parameter_space2}
\end{center}
\end{figure}

Fig.~\ref{fig:parameter_space2} shows the parameter space for the $n=1$ and $n=2$ cases. 
There is no substantial bound from the direct detection experiments for these cases because the non-relativistic elastic cross section is highly suppressed by a small momentum transfer. 
However the IceCube bound still exists as indirect detection because the high energy neutrino is always accompanied by the boosted dark matter in our setup. 
The IceCube bounds for the $n=1$ and $n=2$ cases have been obtained as same as the left panel in Fig.~\ref{fig:parameter_space} by translating the IceCube bound 
for the SD ($n=0$) cross section~\cite{IceCube:2021xzo}. 
The black stars represent the benchmark parameter sets (BP1 and BP2) discussed in the previous section. 

For the SD ($n=1$) case, it can be seen that the numbers of neutrino and boosted dark matter signals can be comparable, 
and the dark matter mass range that the DUNE experiment can test both signals simultaneously is approximately below $8~\mathrm{GeV}$ due to the IceCube bound without the systematic uncertainty. 
For $n=2$, the DUNE sensitivity for the boosted dark matter becomes much higher than the other cases. 
This is expected because the capture rate in the Sun becomes larger than the $n=1$ case due to the gravitational force when the dark matter is captured, 
and there is no strong bound from the direct detection experiments. 

Finally, we comment on a lighter dark matter case ($m_\chi\lesssim 4~\mathrm{GeV}$). 
In this case, the elastic scattering and dark matter semi-annihilation processes may not equilibrate. 
Therefore an extra suppression factor would be multiplied to the capture rate in Eqs.~(\ref{eq:nu_signal}) and (\ref{eq:chi_signal}), and 
the DUNE sensitivity is anticipated to be weaker than the naive extrapolation of the DUNE sensitivity lines in Fig.~\ref{fig:parameter_space} and Fig.~\ref{fig:parameter_space2}.

\section{Summary}
\label{sec:summary}
No dark matter signal at direct detection experiments may infer that the elastic scattering cross section between dark matter and nucleons is suppressed by a small momentum transfer. 
Such kind of dark matter can be explored by the large neutrino volume detectors if the dark matter is boosted by some mechanism. 
In this work, we considered the dark matter semi-annihilation process $\chi\chi\to \nu\overline{\chi}$, which produce the high energy neutrino and boosted dark matter simultaneously. 
These produced particles can be a distinctive signal of the dark matter semi-annihilation. 
In particular, dark matter particles are accumulated in the centre of the Sun, and the semi-annihilation produces the monochromatic neutrino and boosted dark matter. 
We parametrized the differential cross section between dark matter and nucleons for three different cases as in Eq.~(\ref{eq:xsec}) with $n=0,1$ and $2$. 
Then we generated the signal events and atmospheric neutrino background events using \texttt{GENIE}, and estimated the detection sensitivity at DUNE. 
We found that the simultaneous detection of the neutrino and boosted dark matter at DUNE or a combination of DUNE and the other neutrino experiments is possible in some parameter space for $n=1$ and $2$ (momentum transfer dependent cases) 
while $n=0$ (constant case) is completely ruled out by the current direct detection experiments as one can easily expect. 
We found that the dark matter mass region which is testable for the two kinds of signals at DUNE is below $8~\mathrm{GeV}$. 
The larger dark matter mass region is still possible to search by combining DUNE and the other neutrino experiments such as IceCube/DeepCore and Hyper-Kamiokande.

\section*{Acknowledgments}
The authors would like to thank the members of the GENIE working group, especially Costas Andreopoulos, Joshua Barrow, Joshua Berger and Robert Hatcher for helping to use \texttt{GENIE}. 
This work was supported by JSPS KAKENHI Grant Numbers 20H00160, 23K03384 and 23H04004.


\begin{thebibliography}{200}
\bibitem{XENON:2018voc}
E.~Aprile \textit{et al.} [XENON],
Phys. Rev. Lett. \textbf{121}, no.11, 111302 (2018)
[arXiv:1805.12562 [astro-ph.CO]].

\bibitem{PandaX-4T:2021bab}
Y.~Meng \textit{et al.} [PandaX-4T],
Phys. Rev. Lett. \textbf{127}, no.26, 261802 (2021)
[arXiv:2107.13438 [hep-ex]].

\bibitem{LZ:2022ufs}
J.~Aalbers \textit{et al.} [LZ],
[arXiv:2207.03764 [hep-ex]].




\bibitem{Gross:2017dan}
C.~Gross, O.~Lebedev and T.~Toma,
Phys. Rev. Lett. \textbf{119}, no.19, 191801 (2017)
[arXiv:1708.02253 [hep-ph]].


\bibitem{Ipek:2014gua}
S.~Ipek, D.~McKeen and A.~E.~Nelson,
Phys. Rev. D \textbf{90}, no.5, 055021 (2014)
[arXiv:1404.3716 [hep-ph]].


\bibitem{Cline:2023hfw}
J.~M.~Cline, M.~Puel and T.~Toma,
[arXiv:2308.01333 [hep-ph]].


\bibitem{Aoki:2012ub}
M.~Aoki, M.~Duerr, J.~Kubo and H.~Takano,
Phys. Rev. D \textbf{86}, 076015 (2012)
[arXiv:1207.3318 [hep-ph]].

\bibitem{Aoki:2013gzs}
M.~Aoki, J.~Kubo and H.~Takano,
Phys. Rev. D \textbf{87}, no.11, 116001 (2013)
[arXiv:1302.3936 [hep-ph]].

\bibitem{Agashe:2014yua}
K.~Agashe, Y.~Cui, L.~Necib and J.~Thaler,
JCAP \textbf{10}, 062 (2014)
[arXiv:1405.7370 [hep-ph]].

\bibitem{Aoki:2014lha}
M.~Aoki, J.~Kubo and H.~Takano,
Phys. Rev. D \textbf{90}, no.7, 076011 (2014)
[arXiv:1408.1853 [hep-ph]].

\bibitem{Kong:2014mia}
K.~Kong, G.~Mohlabeng and J.~C.~Park,
Phys. Lett. B \textbf{743}, 256-266 (2015)
[arXiv:1411.6632 [hep-ph]].

\bibitem{Kopp:2015bfa}
J.~Kopp, J.~Liu and X.~P.~Wang,
JHEP \textbf{04}, 105 (2015)
[arXiv:1503.02669 [hep-ph]].

\bibitem{Alhazmi:2016qcs}
H.~Alhazmi, K.~Kong, G.~Mohlabeng and J.~C.~Park,
JHEP \textbf{04}, 158 (2017)
[arXiv:1611.09866 [hep-ph]].

\bibitem{Aoki:2017eqn}
M.~Aoki, D.~Kaneko and J.~Kubo,
Front. in Phys. \textbf{5}, 53 (2017)
[arXiv:1711.03765 [hep-ph]].

\bibitem{Aoki:2018gjf}
M.~Aoki and T.~Toma,
JCAP \textbf{10}, 020 (2018)
[arXiv:1806.09154 [hep-ph]].

\bibitem{Kim:2019had}
D.~Kim, J.~C.~Park and S.~Shin,
Phys. Rev. D \textbf{100}, no.3, 035033 (2019)
[arXiv:1903.05087 [hep-ph]].


\bibitem{Berger:2014sqa}
J.~Berger, Y.~Cui and Y.~Zhao,
JCAP \textbf{02}, 005 (2015)
[arXiv:1410.2246 [hep-ph]].

\bibitem{Kelly:2019wow}
K.~J.~Kelly and Y.~Zhang,
Phys. Rev. D \textbf{99}, no.5, 055034 (2019)
[arXiv:1901.01259 [hep-ph]].

\bibitem{Berger:2019ttc}
J.~Berger, Y.~Cui, M.~Graham, L.~Necib, G.~Petrillo, D.~Stocks, Y.~T.~Tsai and Y.~Zhao,
Phys. Rev. D \textbf{103}, no.9, 095012 (2021)
[arXiv:1912.05558 [hep-ph]].

\bibitem{Toma:2021vlw}
T.~Toma,
Phys. Rev. D \textbf{105}, no.4, 043007 (2022)
[arXiv:2109.05911 [hep-ph]].

\bibitem{Guo:2023kqt}
J.~Guo, L.~Wu and B.~Zhu,
Phys. Lett. B \textbf{840}, 137853 (2023)
[arXiv:2302.06159 [hep-ph]].



\bibitem{IceCube:2015rnn}
M.~G.~Aartsen \textit{et al.} [IceCube],
Eur. Phys. J. C \textbf{75}, no.10, 492 (2015)
[arXiv:1505.07259 [astro-ph.HE]].

\bibitem{Super-Kamiokande:2015xms}
K.~Choi \textit{et al.} [Super-Kamiokande],
Phys. Rev. Lett. \textbf{114}, no.14, 141301 (2015)
[arXiv:1503.04858 [hep-ex]].

\bibitem{Hyper-Kamiokande:2018ofw}
K.~Abe \textit{et al.} [Hyper-Kamiokande],
[arXiv:1805.04163 [physics.ins-det]].

\bibitem{DUNE:2020ypp}
B.~Abi \textit{et al.} [DUNE],
[arXiv:2002.03005 [hep-ex]].










\bibitem{Andreopoulos:2009rq}
C.~Andreopoulos, A.~Bell, D.~Bhattacharya, F.~Cavanna, J.~Dobson, S.~Dytman, H.~Gallagher, P.~Guzowski, R.~Hatcher and P.~Kehayias, \textit{et al.}
Nucl. Instrum. Meth. A \textbf{614}, 87-104 (2010)
[arXiv:0905.2517 [hep-ph]].

\bibitem{DUNE:2015lol}
R.~Acciarri \textit{et al.} [DUNE],
[arXiv:1512.06148 [physics.ins-det]].


\bibitem{Ghosh:2020lma}
A.~Ghosh, D.~Ghosh and S.~Mukhopadhyay,
JHEP \textbf{08}, 149 (2020)
[arXiv:2004.07705 [hep-ph]].




\bibitem{Garani:2017jcj}
R.~Garani and S.~Palomares-Ruiz,
JCAP \textbf{05}, 007 (2017)
[arXiv:1702.02768 [hep-ph]].



\bibitem{Ma:2007gq}
E.~Ma,
Phys. Lett. B \textbf{662}, 49-52 (2008)
[arXiv:0708.3371 [hep-ph]].

\bibitem{Aoki:2014cja}
M.~Aoki and T.~Toma,
JCAP \textbf{09}, 016 (2014)
[arXiv:1405.5870 [hep-ph]].

\bibitem{Miyagi:2022gvy}
T.~Miyagi and T.~Toma,
JHEP \textbf{07}, 027 (2022)
[arXiv:2201.05412 [hep-ph]].




\bibitem{Baratella:2013fya}
P.~Baratella, M.~Cirelli, A.~Hektor, J.~Pata, M.~Piibeleht and A.~Strumia,
JCAP \textbf{03}, 053 (2014)
[arXiv:1312.6408 [hep-ph]].


\bibitem{Chauhan:2023zuf}
B.~Chauhan, M.~H.~Reno, C.~Rott and I.~Sarcevic,
[arXiv:2308.16134 [hep-ph]].



\bibitem{Busoni:2017mhe}
G.~Busoni, A.~De Simone, P.~Scott and A.~C.~Vincent,
JCAP \textbf{10}, 037 (2017)
[arXiv:1703.07784 [hep-ph]].


\bibitem{Andreopoulos:2015wxa}
C.~Andreopoulos, C.~Barry, S.~Dytman, H.~Gallagher, T.~Golan, R.~Hatcher, G.~Perdue and J.~Yarba,
[arXiv:1510.05494 [hep-ph]].



\bibitem{ParticleDataGroup:2018ovx}
M.~Tanabashi \textit{et al.} [Particle Data Group],
Phys. Rev. D \textbf{98}, no.3, 030001 (2018)



\bibitem{Super-Kamiokande:2015qek}
E.~Richard \textit{et al.} [Super-Kamiokande],
Phys. Rev. D \textbf{94}, no.5, 052001 (2016)
[arXiv:1510.08127 [hep-ex]].

\bibitem{Honda:2015fha}
M.~Honda, M.~Sajjad Athar, T.~Kajita, K.~Kasahara and S.~Midorikawa,
Phys. Rev. D \textbf{92}, no.2, 023004 (2015)
[arXiv:1502.03916 [astro-ph.HE]].


\bibitem{DUNE:2021gbm}
A.~A.~Abud \textit{et al.} [DUNE],
JCAP \textbf{10}, 065 (2021)
[arXiv:2107.09109 [hep-ex]].

\bibitem{Friedland:2018vry}
A.~Friedland and S.~W.~Li,
Phys. Rev. D \textbf{99}, no.3, 036009 (2019)
[arXiv:1811.06159 [hep-ph]].



\bibitem{Berger:2018urf}
J.~Berger,
[arXiv:1812.05616 [hep-ph]].


\bibitem{Gelmini:2018ogy}
G.~B.~Gelmini, V.~Takhistov and S.~J.~Witte,
JCAP \textbf{07}, 009 (2018)
[erratum: JCAP \textbf{02}, E02 (2019)]
[arXiv:1804.01638 [hep-ph]].


\bibitem{soltrack}
M.~van~der~Sluys, P.~van~Kan, and P.~Sonneveld, AIP Conference Proceedings 1679 (2015), no. 1 080003,
\url{https://aip.scitation.org/doi/10.1063/1.4931544}. 


\bibitem{Rott:2019stu}
C.~Rott, D.~Jeong, J.~Kumar and D.~Yaylali,
JCAP \textbf{07}, 006 (2019)
[arXiv:1903.04175 [astro-ph.HE]].




\bibitem{Kelly:2019itm}
K.~J.~Kelly, P.~A.~Machado, I.~Martinez Soler, S.~J.~Parke and Y.~F.~Perez Gonzalez,
Phys. Rev. Lett. \textbf{123}, no.8, 081801 (2019)
[arXiv:1904.02751 [hep-ph]].

\bibitem{Kelly:2021jfs}
K.~J.~Kelly, P.~A.~N.~Machado, I.~Martinez-Soler and Y.~F.~Perez-Gonzalez,
JHEP \textbf{05}, 187 (2022)
[arXiv:2110.00003 [hep-ph]].

\bibitem{DeRomeri:2021xgy}
V.~De Romeri, P.~Mart\'\i{}nez-Mirav\'e and M.~T\'ortola,
JCAP \textbf{10}, 051 (2021)
[arXiv:2106.05013 [hep-ph]].




\bibitem{PICO:2019vsc}
C.~Amole \textit{et al.} [PICO],
Phys. Rev. D \textbf{100}, no.2, 022001 (2019)
[arXiv:1902.04031 [astro-ph.CO]].

\bibitem{Schumann:2015cpa}
M.~Schumann, L.~Baudis, L.~B\"utikofer, A.~Kish and M.~Selvi,
JCAP \textbf{10}, 016 (2015)
[arXiv:1506.08309 [physics.ins-det]].

\bibitem{DARWIN:2016hyl}
J.~Aalbers \textit{et al.} [DARWIN],
JCAP \textbf{11}, 017 (2016)
[arXiv:1606.07001 [astro-ph.IM]].

\bibitem{ARGO}
GADMC Collaboration, C. Galbiati et al., 
Future Dark Matter Searches with Low-Radioactivity
Argon, Input to the European Particle Physics Strategy Update 2018-2020 (2018).

\bibitem{LZ:2018qzl}
D.~S.~Akerib \textit{et al.} [LZ],
Phys. Rev. D \textbf{101}, no.5, 052002 (2020)
[arXiv:1802.06039 [astro-ph.IM]].

\bibitem{Giroux:2021vpy}
G.~Giroux,
J. Phys. Conf. Ser. \textbf{2156}, 012068 (2021)



\bibitem{IceCube:2021xzo}
R.~Abbasi \textit{et al.} [IceCube],
Phys. Rev. D \textbf{105}, no.6, 062004 (2022)
[arXiv:2111.09970 [astro-ph.HE]].

\end{thebibliography}
\end{document}